# Thermally Activated Processes for Ferromagnet Intercalation in Graphene-Heavy Metal Interfaces

F. Ajejas[1,2], A. Anadon[1], A. Gudin[1], J. M. Diez[1,2], C. G. Ayani[1], P. Olleros[1], L. de Melo Costa[1], C. Navío[1], [1], A. Gutierrez[4], F. Calleja[1], A. L. Vázquez de Parga[1,2,3], R. Miranda[1,2,3], J. Camarero[1,2,3], and P. Perna[1*]

[1] IMDEA Nanociencia, c/ Faraday 9, Campus de Cantoblanco, 28049 Madrid, Spain.

[2] Dpto. Física de la Materia Condensada & Instituto "Nicolás Cabrera", Universidad Autónoma de Madrid, 28049 Madrid, Spain.

[3] IFIMAC, Universidad Autónoma de Madrid, 28049 Madrid, Spain.

[4] Dpto. Física Aplicada & Instituto "Nicolás Cabrera", Universidad Autónoma de Madrid, 28049 Madrid, Spain.

[*]*Corresponding Author: paolo.perna@imdea.org*

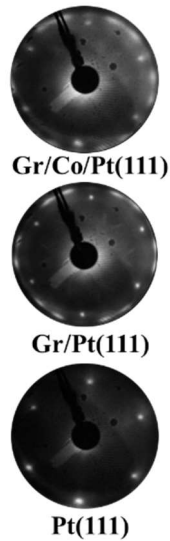
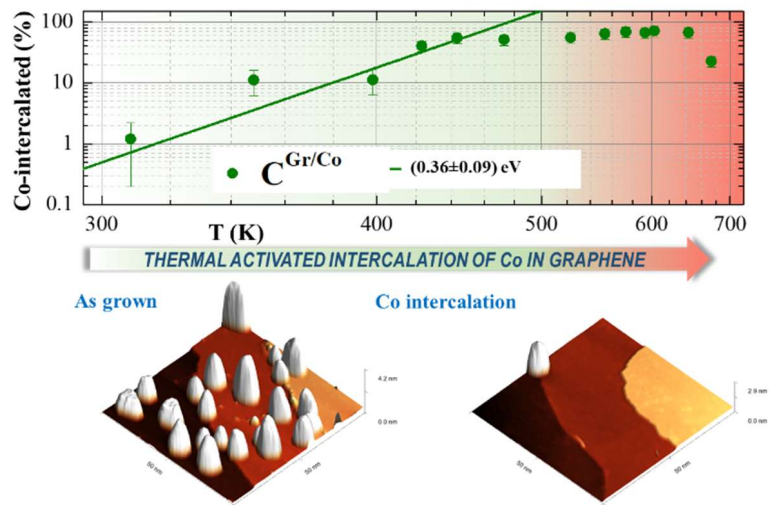

The development of graphene (Gr) spintronics requires the ability to engineer epitaxial Gr heterostructures with interfaces of high quality, in which the intrinsic properties of Gr are modified through proximity with a ferromagnet to allow for efficient room temperature spin manipulation or the stabilization of new magnetic textures. These heterostructures can be prepared in a controlled way by intercalation through graphene of different metals. Using photoelectron spectroscopy (XPS) and Scanning Tunneling Microscopy (STM), we achieve a nanoscale control of thermal activated intercalation of homogeneous ferromagnetic (FM) layer underneath epitaxial Gr grown onto (111)-oriented heavy metal (HM) buffers deposited in turn onto insulating oxide surfaces. XPS and STM demonstrate that Co atoms evaporated on top of Gr arrange in 3D clusters, and, upon thermal annealing, penetrate through and diffuse below Gr in a 2D fashion. The complete intercalation of the metal occurs at specific temperatures depending on the type of metallic buffer. The activation energy and the optimum temperature for the intercalation processes are determined. We describe a reliable method to fabricate and characterize in-situ high quality Gr-FM/HM heterostructures enabling the realization of novel spin-orbitronic devices that exploits the extraordinary properties of Gr.

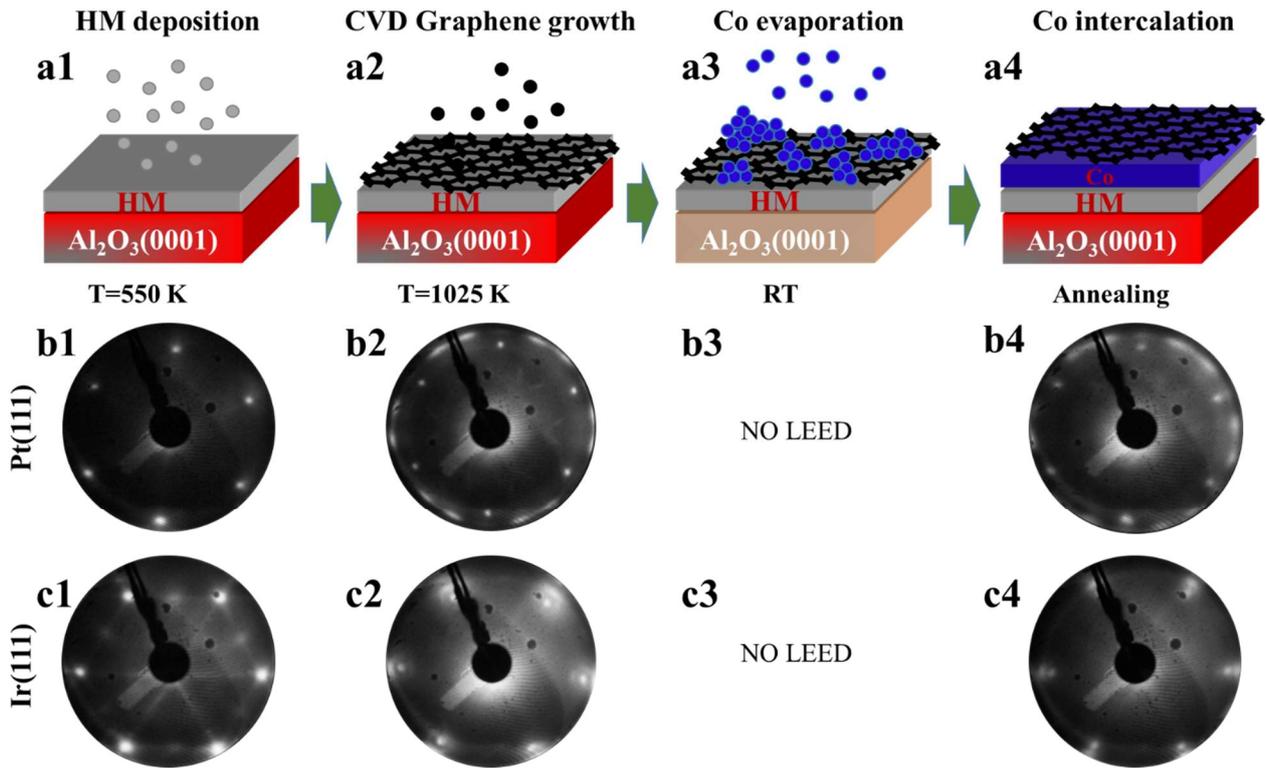

**Figure 1. Sketches of the fabrication of the Gr-based heterostructure and corresponding LEED patterns.** Schematic of the UHV epitaxial growth process: a1) dc sputtering deposition of HM buffer layer (Pt and Ir) at high temperature; a2) Gr growth by CVD; a3) e-beam evaporation of 5 MLs Co at RT on the Gr surface; a4) Co intercalation underneath Gr by thermal annealing. The corresponding LEED patterns (i.e at each stage of the growth) on the Pt and Ir (111)fcc buffers are shown in panels b and c, respectively. In b2 (b4) the hexagonal diffraction spots correspond to the lattice of Pt (Co, pseudomorphic with Pt). The rings around the spots are due the rotational domains of Gr (oriented at ±15º). In c2 (c4) the spots correspond to the hexagonal fcc lattice of Ir (Co, pseudomorphic with Ir). The circles around the spots are the Moiré superstructure of Gr. Note that in the case of Co on top of Gr (panels 3), the lack of long range structural order due to isolated Co clusters is responsible for the absence of the diffraction peaks.

The future sensing and data storage technologies will be based in the exploitation of the spin orbit physics in ferromagnetic-heavy metal systems. Graphene (Gr) is considered an ideal material for room temperature (RT) spintronics because of its unique and intrinsically tuneable electronic properties [1][2][3]. However, its technological development for this purpose relies on the capability to engineer high quality interfaces in which the negligible intrinsic spin-orbit interaction in Gr would be modified to allow for efficient RT spin injection, detection, manipulation or gating.

The electronic properties of Gr can be tuned by proximity with heavier metallic atoms in hybrid Gr/ferromagnet (FM) [4] or Gr/Heavy Metals (HM) [5][6] structures. This may enable manipulating (through spin-orbit effects) the electron's spins for information storage and sensing applications more efficiently, faster and with lower energy consumption than today's commercialized electronics. In particular, when Gr is coupled with a FM, a variety of fundamental and technologically relevant effects have been found, such as high spin injection efficiency [7], antiferromagnetic coupling [8], Rashba effect [5][9], spin filtering [10], tunnel magnetoresistance [11], or enhancement of the magnetic moment perpendicular to the surface [12][13][14]. More recently, it has been demonstrated that Gr/FM structures allow for RT stable chiral spin textures [13][15], and preserve spins over large distances and long times [16].

A convenient method of fabrication of complex Gr/FM/HM heterostructures is by *intercalation*, i.e. by penetration of ferromagnetic atoms through the Gr sheet which, acting as a surfactant [17], will promote an ordered, layer-by-layer, growth below, i.e. at the interface between Gr and the initial substrate. This phenomenon has been observed for graphene in a large number of systems, comprising different substrates and a variety of atomic and molecular species [16]]. Additionally, Gr protects the intercalated species from air, preventing the oxidation of FM layers in complex spintronic devices [13]. There have been several studies to try to identify the mechanisms of intercalation [19][21]. However, to date there are almost no systematic studies that characterize the *activation energies* involved in the intercalation processes.

In this work, we present a complete study of the growth process of epitaxial Gr-based heterostructures, in which a ferromagnetic Co layer is intercalated between Gr and heavy metallic buffer (Pt and Ir) deposited epitaxially on top of insulating $Al_2O_3$ (0001) substrates. The structural, chemical and electronic properties of the surfaces are investigated in-situ by surface sensitive techniques, such as low energy electron diffraction (LEED) and X-ray photoemission spectroscopy (XPS) at each stage of the growth. A close view of the morphology of the surfaces obtained by scanning tunneling microscopy (STM) reveals the specific intercalation processes for Cobalt. The first, which is characterized by low activation temperature and relatively low energy barrier, accounts for the penetration of the FM species through Gr. At intermediate temperatures, the Co adatoms diffuse underneath Gr and form a flat, homogeneous layer that continues growing layer by layer. Higher temperatures promote finally the intermixing of Co with the metallic buffer, which should be avoided since it is, a priori, detrimental for the overall interface properties.

**Growth of Gr/Co on epitaxial HM-buffers onto (111)-oriented insulating oxide surface.**

The samples were prepared entirely in ultra-high-vacuum (UHV) conditions by adopting a specific methodology [13] monitored by in-situ surface analysis, as schematically shown in **Figure 1:** First (panel a1), 30 nm-thick, epitaxial, single-crystal, Pt(111) or Ir(111) buffers were grown by dc sputtering at 550 K on commercial $Al_2O_3$(0001) single crystal substrates. The HM buffers present a crystal quality equivalent to Pt(111) and Ir(111) single crystals, as judged from the corresponding LEED patterns (Fig. 1 b1 and c1). Second (panel a2), epitaxial layers of graphene were grown in-situ by ethylene chemical vapour deposition at 1025 K on top of the buffers. In the case of the Pt (111) buffer, the typical LEED rings (Fig.1 b2) are due to the characteristic [27][28] multi-domain Gr flakes i.e., domains oriented ±15º. The different domains are also revealed by STM on the as-grown samples (see Figure S2 in Supp. Info.). On the contrary, Gr grown on the Ir (111) buffers presents a LEED pattern (Fig.1 c2) corresponding to the well-known "10x10" moiré pattern due to the incommensurate unit cells of Ir and Gr [29].

Third (panel a3), 5 monolayers (MLs)-thick cobalt films were evaporated on top of graphene by molecular beam epitaxy (MBE) at RT with a low deposition rate of 0.3 Å/s monitored by an in-situ quartz balance. Finally (panel a4), the temperature was raised in order to activate the intercalation processes studied by surface microscopy and spectroscopies. After the intercalation process the LEED patterns of the initial Gr/Pt(111) of Gr/Ir(111) surfaces are recovered (Fig. 1 b4 and c4, respectively). This indicates that intercalated Co is pseudomorphic with the metallic substrates, epitaxial and (111)-oriented, as well as homogeneously (over ~µm²) intercalated underneath Gr. Thus, by following this growth methodology, homogeneous epitaxial FCC cobalt layers sandwiched between the HM buffers and Gr were obtained [13].

**Figure 2** shows the XPS spectra of the C *1s* core levels of Gr grown on Pt-(panels a) and Ir- (panels b) buffers before the evaporation of Co and after its intercalation. The C *1s* XPS signals in both structures are clearly visible after the growth of Gr on top of the epitaxial buffers (top panels). The C *1s* line shape is fitted with one main asymmetric component corresponding to the bonds of C atoms of graphene with each HM. In the case of the Pt-buffer (panel a), the C *1s* peak is found at Binding Energy (BE) = 284.0 eV ($C^{Gr-Pt}$, red patterned area), whereas for the Ir-buffer (panel b) the C *1s* peak is at BE = 284.2 eV ($C^{Gr-Ir}$, red patterned area). The energy position of peaks reflect the sp² hybridization of C-C bonding for Gr on weakly interacting Pt(111) and Ir(111) substrates [30]. The C *1s* level in Gr/Pt(111) [Gr/Ir(111)] is shifted to lower binding energy (BE) by -0.26 eV [-0.12 eV] with respect to the one of HOPG [30][31] (vertical grey line in panels a, b) reflecting the doping of Gr on these substrates [30][32].

After RT evaporation of 5 MLs of Co and thermal annealing during 30 minutes at 595 K for Pt and at 665 K for the Ir buffer, Co is completely intercalated underneath the Gr sheet. The XPS spectrum in the bottom panel of Figure 2a shows for the Pt buffer a C *1s* peak centered at BE = 284.5 eV (i.e., $C^{Gr-Co}$, filled green area), i.e. shifted by +0.5 eV towards *higher* BE than the original $C^{Gr-Pt}$. The C *1s* peak in the Ir-buffer structure (bottom panel in Figure 1b) is centered at BE = 284.7 eV (i.e., $C^{Gr-Co}$, filled green area), i.e. shifted by +0.5 eV towards *higher* BE than the original $C^{Gr-Ir}$. Thus, the insertion of Co results in a Gr surface with opposite doping with respect to the Gr/Pt and Gr/Ir structures.

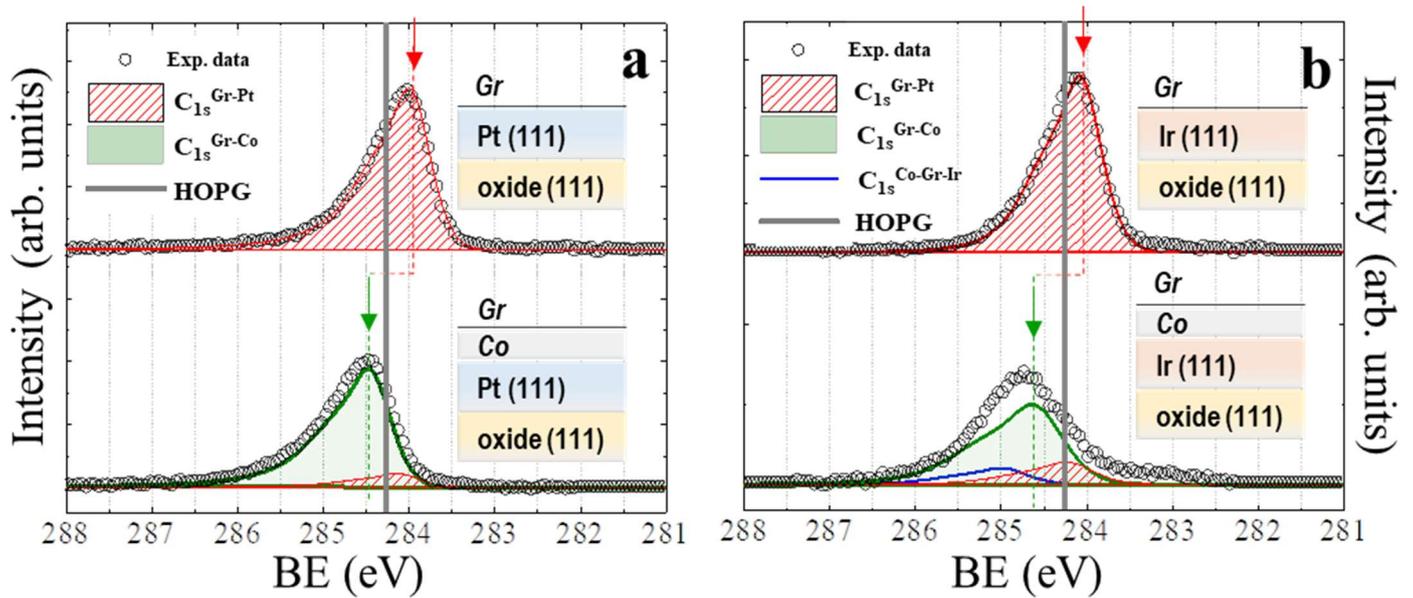

**Figure 2. Analysis of the C 1s XPS core levels for Gr/HM and Gr/Co/HM systems.** (a, top graph): the C 1s peak appears at binding energy (BE) = 284.0 eV (red component in the fit) in Gr/Pt(111). (a, bottom graph): after the Co intercalation [i.e., Gr/Co/Pt(111)], the C *1s* peak is centered at higher energy (284.5 eV) by +0.5 eV. (b, top graph): the C 1s peak appears at binding energy (BE) = 284.2 eV (red component in the fit) in Gr/Ir(111). (b, bottom graph): after the Co intercalation [i.e., Gr/Co/Ir(111)], the C *1s* peak is centered at higher energy (284.7 eV) by +0.5 eV. The vertical grey line indicates the C 1s peak position corresponding to graphite (HOPG). In both cases, 5 MLs of Co were evaporated at RT and completely intercalated by thermal annealing. The sketches of the corresponding structures are indicated at the right of the spectra.

These results are consistent with the predicted values of the BEs of the C *1s* core level of graphene on various metals [31], which depends essentially on the respective metal work functions ($\phi_{Co}$ = 5.0 eV, $\phi_{Pt}$ = 5.93 eV, $\phi_{Ir}$ = 5.37 eV), being Gr p-doped on Pt and Ir and n-doped on Co. The observed shifts in both structures account for the presence of an electric field gradient at the Gr/Co top interface that gives rise to a Rashba-type Dzyaloshinskii-Moriya interaction (DMI) as argued in a previous work [13]. Similar band shift, with Rashba character, was also experimentally found in Gr/Au [5] and Gr/Pb [23][25].

### Co intercalation in Gr/HM-buffer

In order to have a detailed picture of the Co intercalation processes, we have acquired fast (~1 min long) XPS spectra of the C *1s* peaks during in-situ thermal annealing from RT up to ~680 K. The temperature was raised with a rate of ~1 K/min by means of a radiative heater placed under the sample.

Representative *C 1s* spectra as function of the temperature are shown in **Figure 3a** after Co evaporation on the Pt-buffer. The scans in panel a1 and a8 were performed at RT with better statistics and correspond to initial stage of Co/Gr/Pt(111) and to the final, fully intercalated Gr/Co/Pt(111), case respectively, as reproduced in Figure 2.

At each temperature, the C *1s* spectrum is fitted with three components: $C^{Gr-Pt}$ (red patterned area) centered at BE=284.0 eV that corresponds to the Gr bonding to Pt(111) [30]; $C^{Co-Gr-Pt}$ (blue) at 285.0 eV that corresponds to C atoms sandwiched between Co and Pt, i.e. those below the Co islands grown on Gr/Pt(111) and $C^{Gr-Co}$ (green) at 284.5 eV, due to the C atoms interacting exclusively with the intercalated Co.

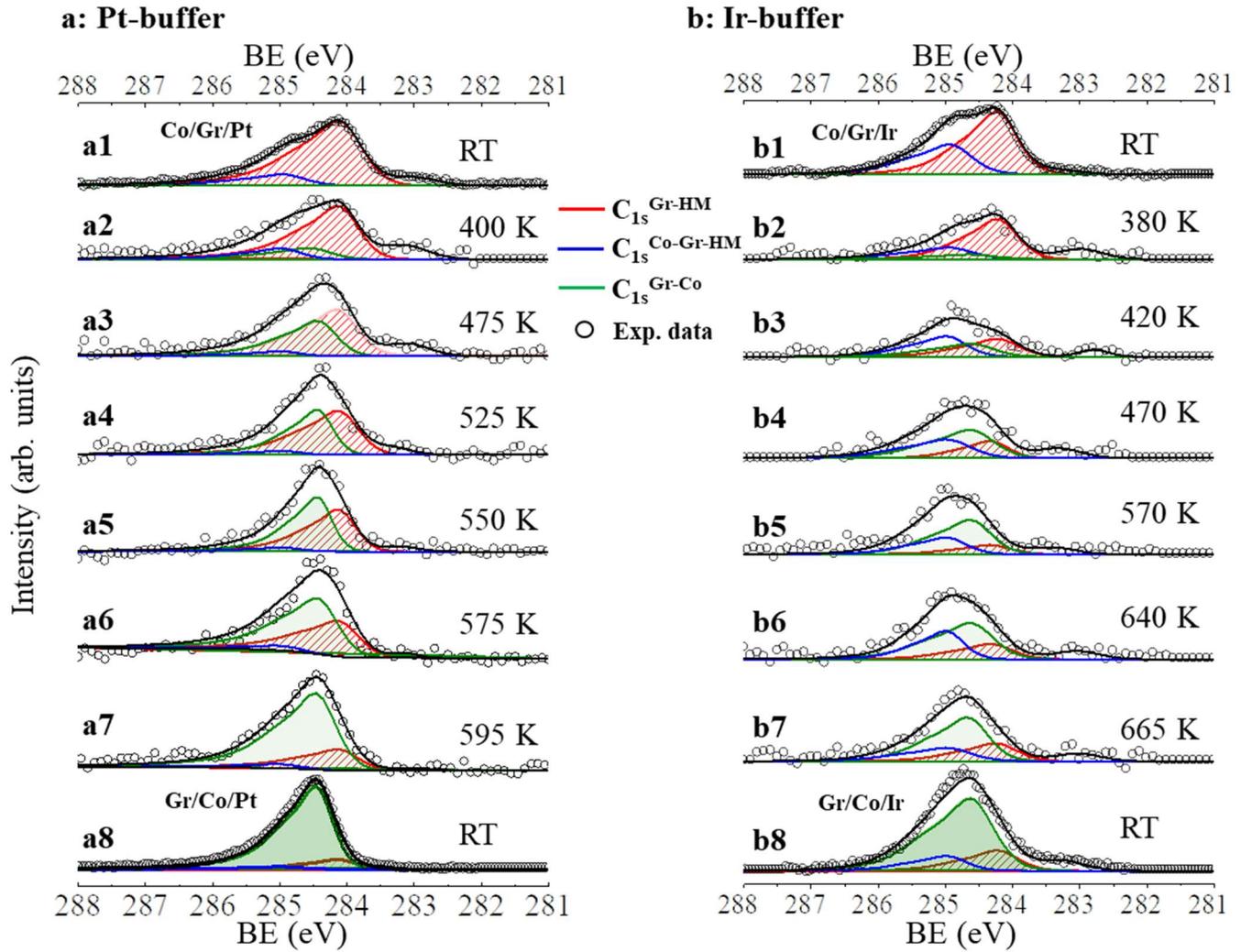

**Figure 3. Temperature evolution of the C 1s XPS spectra.** XPS of the C *1s* region for 5 MLs of Co annealed at different temperatures. (a1- a8): C *1s* spectra for Gr/Pt(111) from RT up to 595 K. (b1-b8): C *1s* spectra of Gr/Ir(111) from RT up to 665 K. The spectra in a8 and b8 were acquired after cooling the sample back to RT. See main text for a detailed description.

If the evaporation of Co would produce a flat Co film covering homogeneously the Gr/Pt(111) substrate prior the annealing procedure, the area of the $C^{Co-Gr-Pt}$ (blue) component in panel a1 should be dominant. Instead, we found that the former (component at 285.0 eV) is only 25% of the total area. This indicates that right after evaporation only a relatively small fraction of the area investigated by XPS (~μm$^2$) is covered by Co, while most of the surface is not covered. This suggests a Volmer-Weber growth mode of Co on top of Gr at RT In fact, the formation of three-dimensional islands is consistent with the absence of LEED patterns in Figure 1. This is further confirmed by STM experiments shown below in Figures 5 and 6.

At 400 K (panel a2), the C *1s* component at BE=284.5 eV due to the C atoms placed on top of Co atoms, i.e. $C^{Gr-Co}$ (green filled area) increases to 13%, indicating a noticeable intercalation, while the $C^{Co-Gr-Pt}$ (blue) component is slightly reduced to about 15% of the total. This indicates that the intercalation occurs at the expense of the area covered by Co clusters. At 475 K (panel a3), the $C^{Co-Gr-Pt}$ component is drastically reduced with a substantial increase in the green intercalated component, which indicates that the Co clusters are disappearing and almost all Co atoms are intercalated. Nevertheless, the $C^{Gr-Pt}$ (red) component is

F. Ajejas *et al.* Thermally activated processes for Ferromagnet intercalation in Graphene-Heavy Metal interfaces. October, 9$^{th}$ 2019.

still >40% of the total area, meaning that a significant fraction of the initial Gr/Pt surface is not yet intercalated. From 525 K (panel a4) on, the $C^{Gr-Pt}$ component starts to decrease while the $C^{Gr-Co}$ slowly increases (panels a5-a7). At 595 K (panel a7), the former is almost totally suppressed, while the latter becomes maximum. The formation of this flat, 2D intercalated Co layer is confirmed by the fact that the LEED pattern observed in Gr/Pt(111) is recovered. By analysing the fraction of the area of each component with respect to the total, we found that more than 75% of the total area of the C *1s* peak corresponds to intercalated Co, i.e. Gr with Co underneath, while the other two components, indicating C atoms still in contact with Pt layer (~20%, non-intercalated area) and with Co on the top (<5%, remaining clusters), are strongly reduced. This quantifies the completeness and uniformity of the intercalation process. Thus, upon intercalation at 595 K, the Co atoms spread underneath Gr to form pseudomorphic MLs in a step-flow like mode [24].

Analogously to the case of Pt buffer, the most representative C *1s* spectra as function of the temperature after the Co evaporation in the Ir-buffer are shown in **Figure 3b**. The scans in panel b1 and b8 were performed at RT with more statistics and correspond to the initial (i.e. Co/Gr/Ir) and final (Gr/Co/Ir) stages of the intercalation process, respectively.

As described above, the C *1s* spectra are fitted at each temperature using three components: $C^{Gr-Ir}$ (red patterned area) at BE=284.2 eV (in agreement with [20][30][32]); $C^{Gr-Co}$ (green) at 284.7 eV due to the C atoms interacting exclusively with the intercalated Co; and, $C^{Co-Gr-Ir}$ (blue) at 285.0 eV that corresponds to C atoms sandwiched between Co and Ir.

After the Co evaporation and prior the annealing (panel b1) a large $C^{Co-Gr-Ir}$ (blue) component is found, unlike the case of the Pt-buffer. This suggests a coverage of the graphene surface by Co clusters *larger* than in the case of Gr/Pt(111). At 420 K (panel b3), the increase in the intensity of the C *1s* $C^{Gr-Co}$ component (green) at BE=284.7 eV indicates a substantial intercalation. By increasing the temperature below 600 K (panels b4-b5), the area of this component increases, indicating the formation of a continuous Co layer below Gr, while $C^{Gr-Ir}$ decreases to less than 17% of the total. This behaviour is analogous to the case of the Pt-buffer. By analysing the fraction of the area of each component with respect to the total, we found that ~60% of the total area of the C *1s* peak corresponds to intercalated Co. If the temperature is further increased (panels b6-b7), we notice the slight increase of the $C^{Gr-Ir}$ (red line) evidencing the intermixing of intercalated Co into the Ir buffer.

XPS spectra at each stage of growth in the region of the Pt *4d*, Ir *4d* and Co *2p* core levels for Pt and Ir buffers are shown in the Supporting Information (Figure S1). Note that the analyses of the area and position of the Pt and Ir *4d* core levels as well as the Co *2p* levels during the intercalation processes indicates the absence of Co/HM intermixing at temperatures below 650K.

**Thermally activated mechanisms for Co intercalation**

The intercalation processes of different atomic species through epitaxial Gr have been interpreted in terms of diffusion of atoms through Gr island edges [18], wrinkles and defects [21], or direct exchange with carbon atoms [18][19]. The energy required to activate each of these mechanisms is experimentally delivered by thermal annealing, and depends on the specific substrate onto which the Gr is grown and on the atomic species to be intercalated. DFT calculations indicate that the presence of a substrate lowers the energy barrier for the intercalation compared to free standing Gr, and that adatoms on top of Gr further reduce such barrier [22][26].

In weakly interacting metallic substrates, such as Gr/Ir(111), Decker *et al.* [14] found that the intercalation proceeds mainly via migration of atoms through Gr patch edges. Instead, the superior surface quality of highly interacting Gr/Ru(0001) interfaces [33][34], with few pre-existing defects, makes energetically less favoured the penetration through edges or wrinkles [19], while the cooperative effect of the presence of the substrate and Co clusters favours the formation of surface defects enabling hence the direct penetration through the Gr sheet [26].

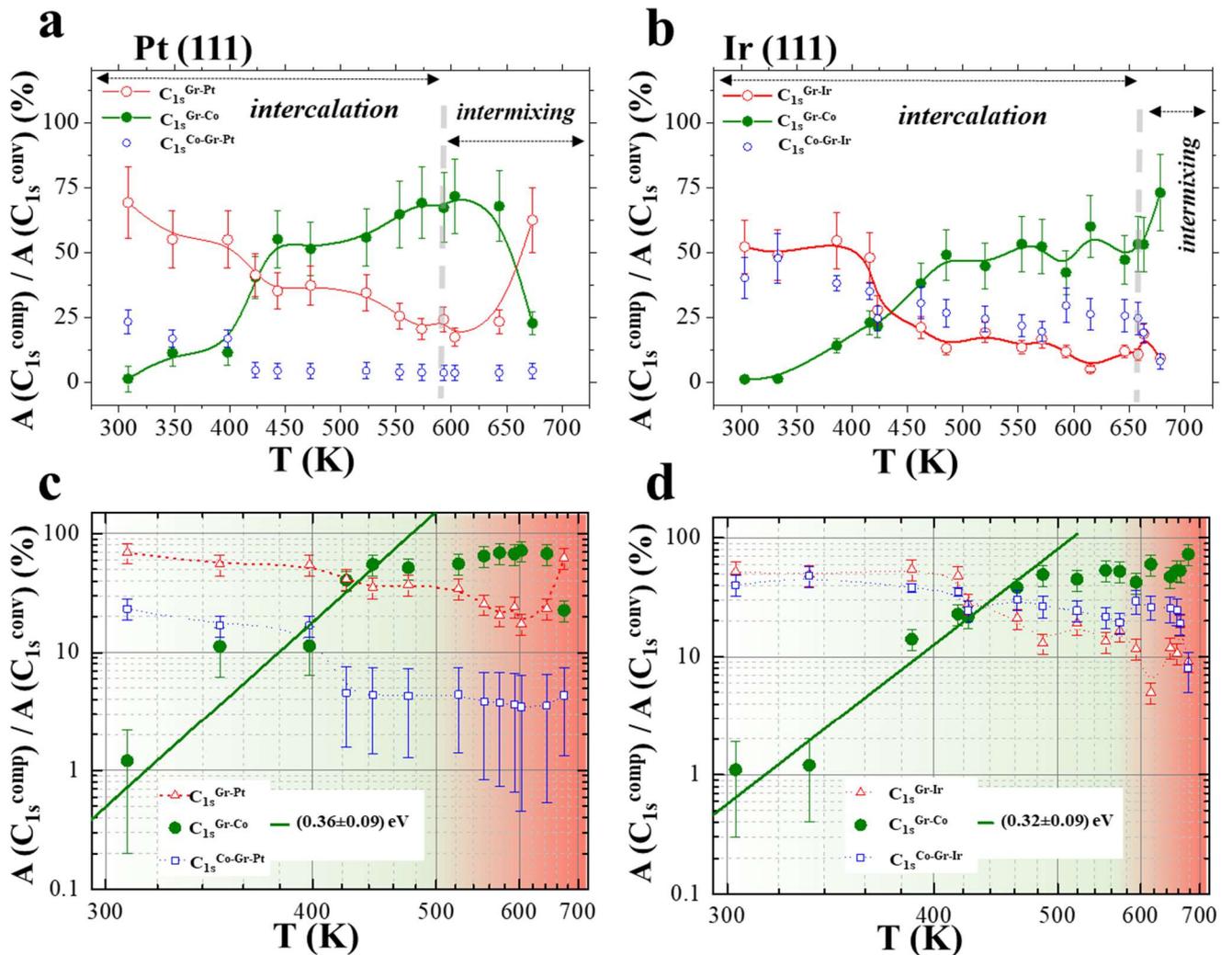

**Figure 4**. **Monitoring of Co intercalation in Gr/Pt and Gr/Ir. a,b)** Temperature evolution of area percentage of the $C^{Gr-Co}$ (green symbols), $C^{Co-Gr-HM}$ (blue) and $C^{Gr-HM}$ (red) components ($A_{C1s}^{comp}$) normalized by the total area of the C $1s$ peak ($A_{C1s}^{conv}$), as taken by the XPS spectra. Two temperature ranges, indicated by grey vertical lines, are identified. In the low temperature regions, i.e. from 300K to 595K for Pt, and from 300K to 645K for Ir, Co atoms penetrate and then diffuse underneath Gr respectively. Above 600 K (for Pt buffer in panel a) and 645 K (for Ir, panel b) the increase of $C^{Gr-HM}$ and decrease of $C^{Gr-Co}$ correspond to a scenario in which Co atoms diffuse into the HM layer while HM atoms move towards the surface. Panels **c** and **d** present in Log scale the ratio $A = A_{C1s}^{comp}/A_{C1s}^{conv}$ as function of T in reciprocal scale. The linear fits (green continuous lines) of $C^{Gr/Co}$ component give the activation energies of the intercalation process.

**Figure 4** shows the area of each component in the C $1s$ XPS spectra ($A_{C1s}^{comp}$) as function of the temperature. These areas (divided by the total area, $A_{C1s}^{conv}$) account for the amount of Co atoms in the different structures. In Figure 4 we can distinguish two temperature regions. The first, from RT to 595 K and 645 K for Pt and Ir respectively, corresponds to the Co intercalation, while the high temperature region corresponds to the Co/HM intermixing.

In the case of Pt buffer (panel a), the first region, from RT to 445 K, is characterized by an exponential increase of the $C^{Gr-Co}$ component area (green symbols) and by a reduction of the $C^{Gr-Pt}$ (red). Both components do not substantially change from 475 K up to 595 K, displaying an almost saturated behaviour. The $C^{Gr-Co}$ reaches about 75% of the total convoluted area while the $C^{Gr-Pt}$ drops to less than 20% and the $C^{Co-Gr-Pt}$ reduces to <5% at 600 K. Finally, the annealing above 650 K causes the increase of the $C^{Gr-Pt}$ and the simultaneous drop of the $C^{Gr-Co}$, indicating that Pt atoms move towards the surface while Co atoms diffuse into the Pt layer (intermixing).

In order to determine the average activation energy, $E_a$, of the intercalation process, we have assumed that $A = A_{C1s}^{comp}/A_{C1s}^{conv}$ can be written as $A(T) \propto e^{-\frac{E_a}{k_B T}}$, where $k_B$ is the Boltzmann constant. From the corresponding plot as function of $1/k_B T$ shown in Figure 4c, an activation energy for the process of Co intercalation underneath Gr/Pt(111) of $E_a$ (Pt)= 0.36 ± 0.09 eV can be obtained.

A similar behaviour is also found in the case of Ir buffer (panel b). However, we can appreciate easily some differences. First, in Gr/Ir(111) about 50% of the evaporated sample area is initially covered by Co clusters; second, the intercalation requires a slightly higher temperature than in Gr/Pt; and third, at the end of the process only 60% of the evaporated Co has been intercalated. The process requires an activation energy comparable to the one found in the Pt case, i.e. $E_a$ (Ir) = 0.32 ± 0.09 eV. These results evidence a somewhat less efficient intercalation mechanism through the Gr/Ir surface than on Gr/Pt in agreement with the larger interaction between Gr and Ir [29] compared to Pt [27][28] and the lack of domain edges indicated by the LEED pattern of Gr/Ir(111) and Gr/Co/Ir(111), which shows a moiré pattern without rotational domains. Notice that the intermixing and alloying between Co and Ir, as indicated by the increase of the $C^{Gr-Ir}$ component, is detected around 650 K, while the intercalation of Co atoms is still proceeding ($C^{Gr-Co}$ increases).

In order to confirm the above picture, an additional STM study was conducted under UHV and in similar conditions to the XPS study for the case of the Pt(111) buffer. First, the Co-free surface of Gr/Pt/oxide(111) revealed atomically flat and large Gr flakes with a large variety of rotational domains, as shown in Figure S2 of the Supporting Information, and in agreement with the LEED characterizations discussed above (Figure 1). Then, different amounts of Co were evaporated at RT and the sample was annealed at various temperatures while checking the surface evolution by means of STM imaging at RT (**Figure 5**).

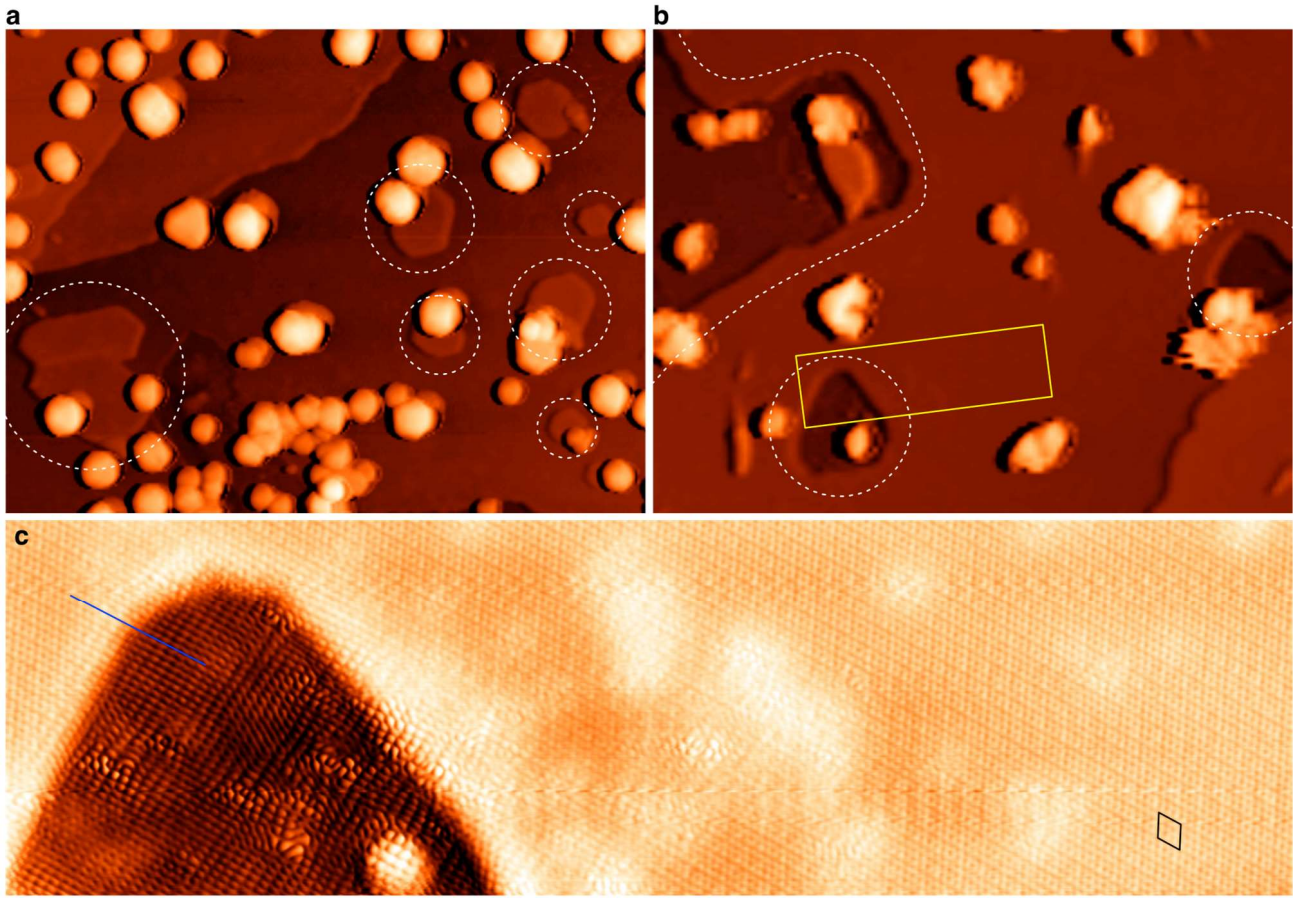

**Figure 5. Initial stages of intercalation.** Panels a and b show 100 nm-wide STM images recorded at +1V bias and 200 pA tunneling current on the early (panel a) and final (panel b) stages of the intercalation of the first ML of Co at 500 K and 900 K. Interface regions between the non-intercalated graphene and the first Co interlayer are highlighted in white dash. Panel c presents a 35 nm-wide STM image with atomic resolution recorded at +30 mV bias and 3 nA tunneling current on the area marked by the yellow rectangle on panel b. The black rhombus highlights a 4x4 moiré pattern in the intercalated region, and the blue line marks a high symmetry direction of the graphene lattice across the interface between the intercalated and non-intercalated regions.

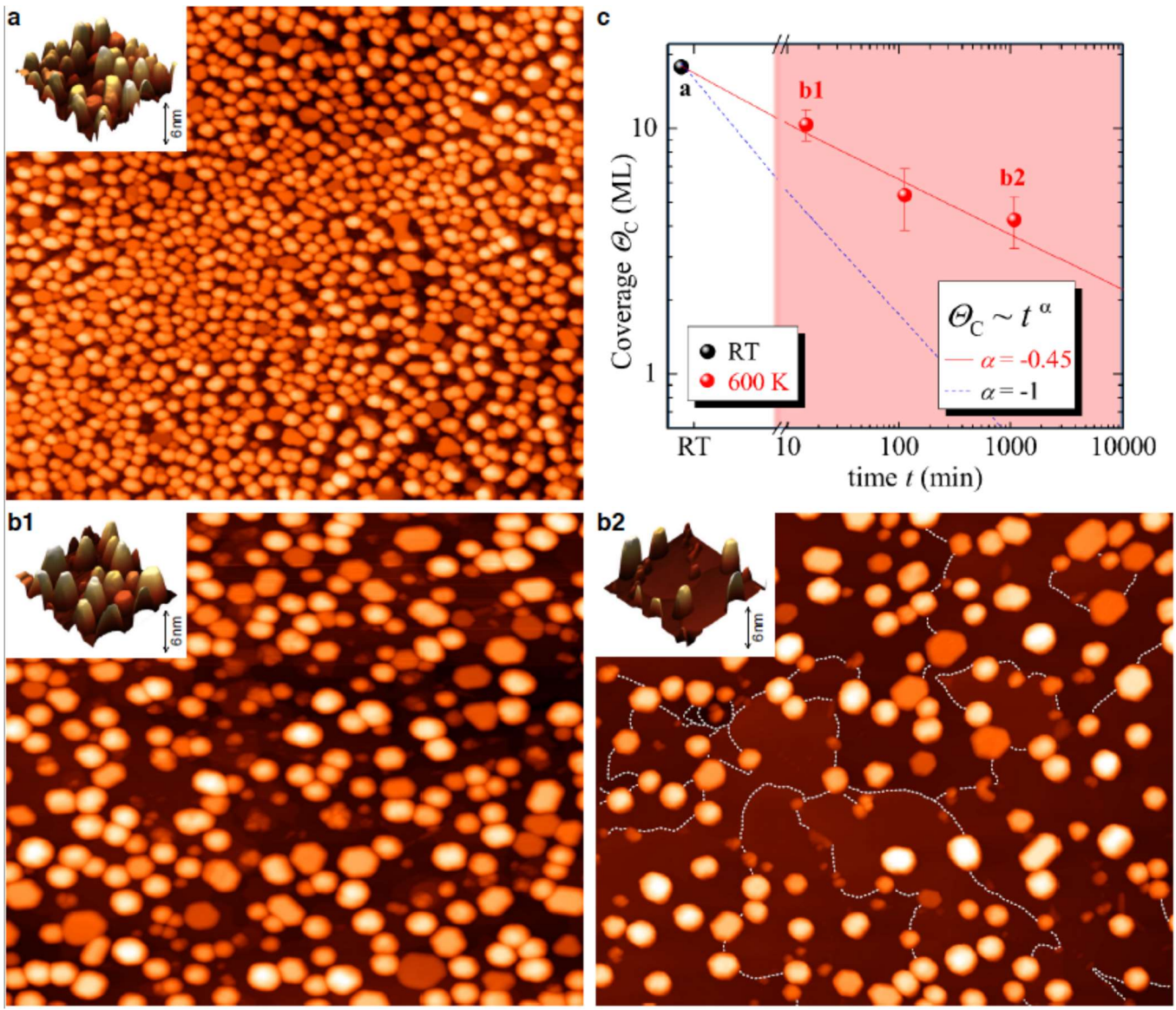

**Figure 6. Intercalation dynamics of 18ML Co studied by STM.** 250 nm side STM images recorded at +1 V bias and 200 pA tunneling current on the non-annealed surface (panel a), annealed to 600 K for 10 minutes (panel b1) and kept at 600 K for 1000 minutes (panel b2). The corresponding insets show 50 nm wide cuts represented in 3D. In panel b2, the monoatomic steps in between Co clusters have been highlighted with white dashed lines. All images have been recorded at RT. Panel c shows the time evolution of the coverage of Co clusters, obtained from the volume analysis of the complete STM annealing sequence (see Figure S3 in Supporting Information).

We first studied the initial stages of intercalation by preparing a low coverage sample, with a nominal Co coverage of 2 MLs. Upon evaporation at RT, Co forms 3D clusters covering only a small fraction of the Gr surface, as deduced above from the XPS results. Annealing the sample to 500 K for 10 minutes results in the STM image shown in Figure 5a. Some intercalated regions are indicated by the white dashed circles showing that, in the initial stages, Co penetrates through Gr *below* the clusters, suggesting a direct Co-C exchange mechanism. Further annealing to 900 K resulted in the completion of the first Co interlayer, as illustrated in Figure 5b.

As a further proof of the intercalation, an atomic resolution STM image corresponding to the region highlighted in yellow in Figure 5b is presented in panel c. The atomic lattice of Gr is continuously resolved both in the intercalated (right) and in the non-intercalated (left) areas, along by the blue line. Also note that a quasi- 4x4 moiré pattern, arising from the mismatch between graphene and the underlying surface is clearly detected in the intercalated area (unit cell marked in black in the lower right corner). This pattern is not resolved in the non-intercalated area due to the presence of a strong inter-valley scattering signal. Note that the temperature for intercalation at low Co coverages is higher compared to the larger coverages discussed previously and in the following.

**Figure 6** shows the time evolution at a constant annealing temperature of 600 K of an initial Co coverage of 18 ML. In this case, similar deposition rates for Co and annealing temperatures to the XPS study were used. The as-deposited surface (panel a) presents a large array of Co clusters, with an average height of 4.7 nm and an average radius of 5.0 nm, in agreement with the 3D growth of Co on Gr at RT previously discussed. As the sample is annealed, the cluster density decreases, exposing gradually the intercalated Gr surface (panels b1 and b2). The total Co cluster coverage $\Theta_C$, obtained from the cluster volume in the STM images, is shown in panel d for the complete annealing sequence. The cluster coverage evolves from 18 ML, i.e., before any annealing, to near 10 ML after 10 min 600K annealing (panel b1), decreasing to 4 ML after the last annealing (1000 min, panel b2). The presence of monoatomic steps in between the remaining Co clusters in the annealed surface (best resolved in panel c, where they are highlighted in white dash) is a clear indication that the intercalation process is mainly 2D, in agreement with the previous analysis of the XPS data. Panel c provides additional insights onto the dynamics of the intercalation processes. For instance, the clusters coverage decreases with annealing time as $\Theta_C \propto t^\alpha$, with a slope of α ~ -0.45±0.05, indicating that the intercalation process follows a universal random walk (Brownian motion) diffusion process.

In summary, the combined study of XPS and STM reveals that *i)* Co atoms evaporated on top of Gr agglomerate to form 3D clusters, *ii)* the intercalation is activated thermally at a specific temperature. Upon thermal annealing, Co atoms penetrate directly through Gr below the clusters. Subsequently, Co starts to diffuse underneath Gr and form a homogeneous layer. For much higher temperature, Co and HM atoms intermix. The first two processes define the complete intercalation of the metal underneath Gr, and follow a behaviour that can be characterized by an activation energy of the order of 0.3 eV, as clearly identified in panel a of Figure 4.

One should note that an incomplete Co intercalation leads to depressed interface and magnetic properties due to *i)* the oxidation of the FM not protected by Gr, *ii)* undefined electronic band gap of Gr, *iii)* negligible effective interfacial DMI. Our in-situ analysis allows controlling the surface properties of each layer and avoiding incomplete intercalation and alloy formation, which are mainly caused by under and over thermal annealing, respectively.

**Conclusions**

In this work, we have introduced a novel methodology based on the combined use of ultra-high vacuum growth, metal intercalation, spectroscopy and tunneling microscopy to fabricate and characterize in-situ high quality Gr-based, perpendicular magnetic anisotropy systems. By combining XPS and STM experiments as function of the annealing temperature, we have discerned the processes of Co intercalation underneath Gr grown epitaxially on two different heavy metal surfaces, Pt and Ir, which were deposited onto (111)-oriented insulating oxides. The thermally activated Co intercalation occurs in two steps in which the deposited atoms penetrate in Gr either via direct Co-C exchange or through surface defects and edges. The intercalated atoms diffuse then in 2D underneath Gr and form flat, homogeneous and crystalline layers. These processes are characterized by a low activation temperature. The Co intercalation occurs at lower temperature and is more efficient in Gr/Pt interfaces than Gr/Ir. Higher temperatures promote the intermixing between the FM and the metallic buffer. We have hence demonstrated that the metal intercalation allows for efficient tuning of the structural and electronic properties of Gr and enables the growth of

## Acknowledgments

This research was supported by the Regional Government of Madrid through Project P2018/NMT-4321 (NANOMAGCOST-CM) and P2018/NMT-4511 (NMAT2D), and by the Spanish Ministry of Economy and Competitiveness (MINECO) through Projects RTI2018-097895-B-C42, FIS2016-78591-C3-1-R, PGC2018-098613-B-C21, PGC2018-093291-B-I00, FIS2015-67367-C2-1-P and PCIN-2015-111 (FLAGERA JTC2015 Graphene Flagship "SOgraph"). IFIMAC acknowledges support from the "Maria de Maeztu" programme for units of Excellence in R&D (MDM-2014-0377). IMDEA Nanoscience is supported by the 'Severo Ochoa' Programme for Centres of Excellence in R&D, MINECO [grant number SEV-2016-0686].

## Contributions

P.P. conceived and designed the project with assistance from F.A., J.C. and R.M. F.A., A.A., A.G. contributed equally to this work realizing the growth, XPS-UPS, LEED and magnetometry experiments with the help of J.M.D., P.O., L.M.C., C.N. and A.G. F.C., C.G.A. and A.V.P. performed the STM experiments. F.A., A.A. F.C. and P.P. treated and analysed the data. P.P. prepared the manuscript with the help of F.A., A.A., F.C., J.C. and R.M. All authors discussed and commented the manuscript.

## Methods

**Growth.** The substrates used in the experiments were 10 mm thick, 5 x 5 mm$^2$ pre-polished $Al_2O_3$ (0001) single crystals. Figure 1 shows the scheme of the growth process. The substrate was placed in a vacuum chamber with base pressure of ~$10^{-8}$ mbar and annealed at 700 K during 1 hour prior the deposition. The epitaxial growth of the buffer layer (panel a1 in Figure 1) was performed by dc sputtering $Ar^+$ deposition with partial gas pressure of $10^{-3}$ mbar at 15 W magnetron power, with parallel configuration between sample and target whose distance was set to 5 cm. The deposition rate was measured by in-situ quartz balance. After the deposition, the sample was cooled down to RT. Two different buffer HM layers [Pt(111) and Ir(111)] have been grown. In-situ LEED measurements confirm the epitaxial growth of both buffers and the FCC structure (ABC stacking sequence). Gr was then grown in-situ by ethylene ($C_2H_4$) dissociation (partial pressure 2 x $10^{-8}$ mbar during 30 minutes) at 1025 K during 1 hour in UHV (base pressure of $10^{-9}$ mbar). The sample was then cooled to RT and the FM layer was evaporated by molecular beam epitaxy (MBE) at 0.04 Å/s. The sample was gradually heated up while acquiring XPS spectra for the real time study of the intercalation process.

**Photoemission experiments.** At each stage of the growth process, we have performed careful LEED and XPS measurements. The XPS measurements were performed by using a monochromatized X-ray source and a hemispherical energy analyzer (SPHERA-U7). The Al Kα line (hν = 1486.7 eV) was used from an Al anode for X-ray Photoelectron Spectroscopy (XPS). The analyser pass energy was set to 20 eV for the XPS measurements to have a resolution of 0.6 eV. The angular acceptance for the used aperture size is defined solely by the magnification mode, i.e. 1750 × 2750 μm$^2$. The core level spectra were fitted to mixed Gaussian-Lorentzian components; all the binding energies are referred to the sample Fermi level. In the fits presented in Figure 3, an additional component centered at 283.2 eV was introduced accounting for leftovers of ethylene dissociation ($C^{C-H}$). This component was only needed in the first low temperature spectra of the thermal annealing.

**Scanning Tunneling Microscopy.** STM measurements have been performed in an independent UHV chamber equipped with an Omicron variable temperature STM (VT-STM 25). The STM head was operated with a Nanonis electronics. Bias voltages given in the manuscript refer to the sample voltage. Gr/Pt/oxide samples were prepared ex-situ, as explained in Growth section, transported to the STM chamber in air and subsequently annealed in-situ to remove air contamination. Co evaporation and the annealing sequences leading to the intercalation process were performed in-situ, as explained in the main text. All STM measurements have been performed at room temperature.